\documentclass[journal=nalefd,manuscript=article]{achemso}

\usepackage{graphicx}
\usepackage{amsmath}
\usepackage{mathtools}
\usepackage{comment}
\usepackage{bbm}
\usepackage{lipsum}
\usepackage{amssymb}
\usepackage{xcolor}
\definecolor{red}{rgb}{1,0.,0}

\usepackage[utf8]{inputenc}
\usepackage{verbatim}
\usepackage{color,soul}
\usepackage{graphicx}% Include figure files
\usepackage{subcaption}
\usepackage{dcolumn}% Align table columns on decimal point
\usepackage{bm}% bold math

\author{Yueh-Chun Wu}
\author{G\'abor B. Hal\'asz}
\affiliation{Materials Science and Technology Division, Oak Ridge National Laboratory, 1 Bethel Valley Rd, Oak Ridge, TN 37831} 
\author{Joshua T. Damron}
\affiliation{Chemical Sciences Division, Oak Ridge National Laboratory, 1 Bethel Valley Rd, Oak Ridge, TN 37831} 
\author{Zheng Gai}
\author{Huan Zhao}
\affiliation{Center for Nanophase Materials Sciences, Oak Ridge National Laboratory, 1 Bethel Valley Rd, Oak Ridge, TN 37831} 
\author{Yuxin Sun}
\affiliation{Department of Physics, Purdue University, West Lafayette, Indiana 47907, USA} 
\altaffiliation{Purdue Quantum Science and Engineering Institute, West Lafayette, IN 47907, USA}
\author{Karin A Dahmen}
\affiliation{Department of Physics, University of Illinois, Urbana-Champaign, IL, 61801, USA} 
\author{Changhee Sohn}
\affiliation{Materials Science and Technology Division, Oak Ridge National Laboratory, 1 Bethel Valley Rd, Oak Ridge, TN 37831}
\altaffiliation{Department of Physics, Ulsan National Institute of Science and Technology, Ulsan 44919, South Korea} 
\author{Erica W. Carlson}
\affiliation{Department of Physics, Purdue University, West Lafayette, Indiana 47907, USA} 
\altaffiliation{Purdue Quantum Science and Engineering Institute, West Lafayette, IN 47907, USA}
\author{Chengyun Hua}
\author{Shan Lin}
\author{Jeongkeun Song}
\author{Ho Nyung Lee}
\author{Benjamin J. Lawrie}
\affiliation{Materials Science and Technology Division, Oak Ridge National Laboratory, 1 Bethel Valley Rd, Oak Ridge, TN 37831} 
\email{lawriebj@ornl.gov}

\title{Nanoscale magnetic ordering dynamics in a high Curie temperature ferromagnet}


\date{\today}% It is always \today, today,
\begin{document}

%\begin{tocentry}

%\begin{figure}
%   \includegraphics[width=2in]{tocfig.png}
%\end{figure}

%\end{tocentry}
\clearpage

\begin{abstract}
Thermally driven transitions between ferromagnetic and paramagnetic phases are characterized by critical behavior with divergent susceptibilities, long-range correlations, and spin dynamics that can span kHz to GHz scales as the material approaches the critical temperature $\mathrm{T_c}$, but it has proven technically challenging to probe the relevant length and time scales with most conventional measurement techniques. In this study, we employ scanning nitrogen-vacancy center based magnetometry and relaxometry to reveal the critical behavior of a high-$\mathrm{T_c}$ ferromagnetic oxide near its Curie temperature. Cluster analysis of the measured temperature-dependent nanoscale magnetic textures points to a 3D universality class with a correlation length that diverges near $\mathrm{T_c}$. Meanwhile, the temperature-dependent spin dynamics, measured through all optical relaxometry suggest that the phase transition is in the XY universality class. Our results capture both static and dynamic aspects of critical behavior, providing insights into universal properties that govern phase transitions in magnetic materials.

\end{abstract}

\section{Introduction}
Understanding continuous phase transitions near critical points is crucial in fundamental physics. In condensed matter physics and materials science, these transitions, often driven by variations in temperature, pressure, and external fields, reveal complex interactions among lattice structures, electronic states, and magnetic moments. Of particular interest are second-order (continuous) phase transitions, where the order parameter changes smoothly and is often associated with spontaneous symmetry breaking. Near a second-order phase transition, systems exhibit critical phenomena, including divergent correlation lengths, fluctuations, and susceptibilities. 
These critical phenomena fall into distinct universality classes set by, {\em e.g.}, the spatial dimension and the symmetry character of the order parameter, such that scaling laws near criticality are independent of microscopic details.
Understanding these behaviors provides insights into the underlying physics that governs a wide range of systems. 

Over the past decade, nitrogen-vacancy (NV) centers in diamond have emerged as powerful nanoscale quantum sensors capable of probing ferromagnetic \cite{zaper2024scanning,celano2021probing,sun2021magnetic} and antiferromagnetic \cite{meisenheimer2024switching,ding2023observation} ordering, and slow telegraph switching of magnetic domain walls has been observed through time-series fluorescence measurements and through variations in the measured optically-detected-magnetic-resonance (ODMR) linewidth\cite{jenkins2019single}. On the other hand, NV $T_1$ and $T_2$  measurements have increasingly been used to measure high-frequency fluctuating magnetic fields that are hard to access with conventional nanoscale magnetic probes like magnetic force microscopy \cite{schafer2014tracking,mccullian2020broadband,du2017control,finco2021imaging,kolkowitz2015probing,andersen2019electron,mclaughlin2022quantum,pelliccione2014two,schmid2015relaxometry,tetienne2016scanning,mccullian2020detection,ziffer2024quantum}. The non-invasive visualization of magnetic textures and spin dynamics in response to changes in temperature or external fields is essential to characterizing critical phenomena associated with continuous phase transitions in magnetic systems where correlation scales diverge in time and space near the critical point. Recent studies have pointed to the potential for spin qubit probes of critical dynamics in spin systems using spin coherence measurements\cite{ziffer2024quantum}, but all-optical $T_1$ probes of critical dynamics would unlock the potential for probing materials in high-field environments where $T_2$ measurements are challenging to implement. 

\begin{figure*}[ht]
\centering
    \includegraphics[width=\textwidth]{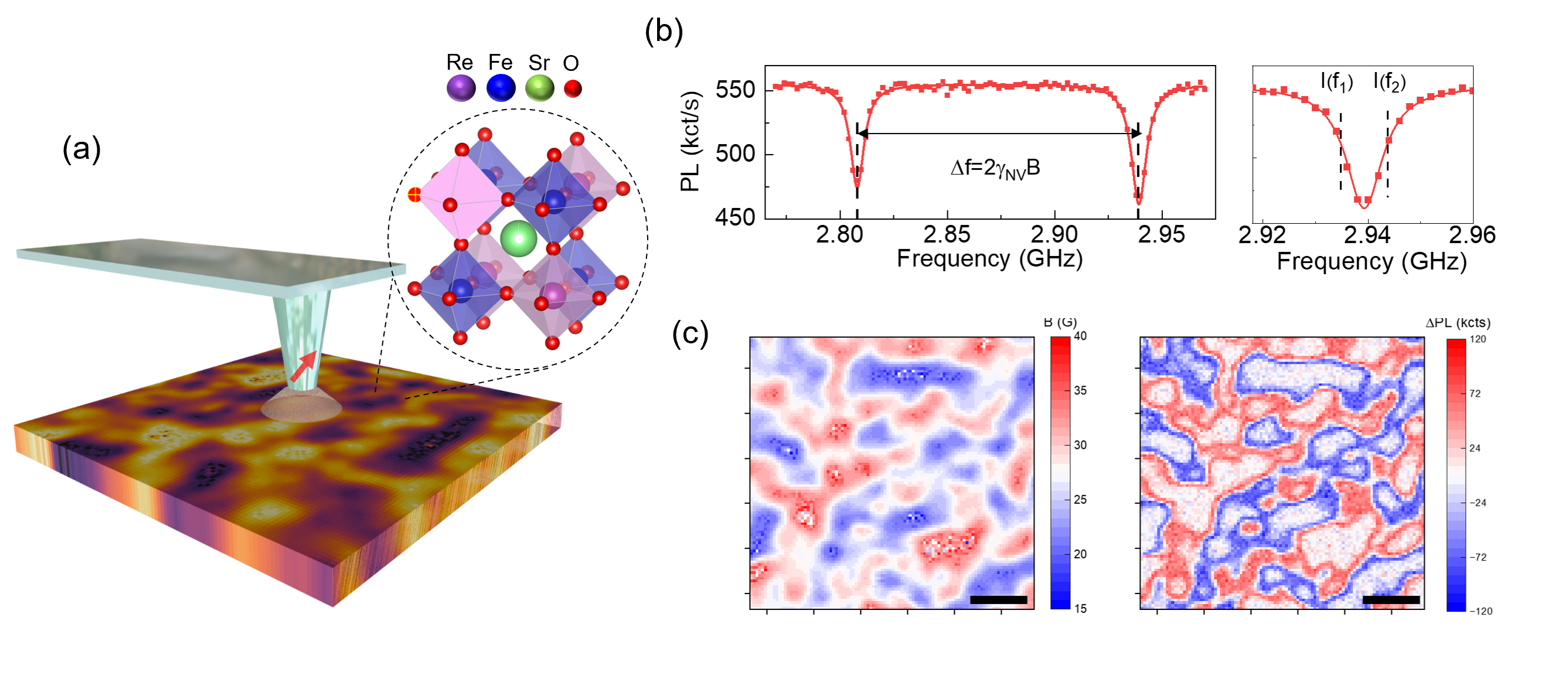}
    \captionsetup{justification=raggedright}
    \caption{(a) Schematic illustration of the NV scanning microscope platform and crystal structure of the high-$\mathrm{T_c}$ ferromagnetic oxide $\mathrm{Sr_2FeReO_6}$. (c) Measured ODMR spectra highlighting (left) the spectrum acquired in full-B mode to quantitatively measure the local magnetic field with no additional calibrations or assumptions and (right) the two frequencies that are used in dual-iso-B mode for high speed imaging of local magnetic textures. (d) Surface magnetic texture of $\mathrm{Sr_2FeReO_6}$ measured at room temperature using (left) full-B mode and (right) dual-iso-B mode. Scale bar: 400nm.
}
    \label{fig:fig1}
\end{figure*}

In this study, we investigate nanoscale magnetic ordering and spin fluctuations near the ferromagnetic-paramagnetic (FM-PM) phase transition in a thin film of the high-$\mathrm{T_c}$ double perovskite strontium iron rhenium oxide ($\mathrm{Sr_2FeReO_6}$) using scanning NV magnetometry and relaxometry. Our results reveal the onset of magnetization characterized by surface magnetic textures that exhibit robust scaling behavior indicative of a second-order phase transition. Concurrently, we observe a pronounced increase in spin fluctuations near $\mathrm{T_c}$, which are detected through NV $\mathrm{T_1}$ relaxometry and modeled using Landau-Ginzburg theory. The combined results highlight the pivotal role of spin dimensionality in $\mathrm{Sr_2FeReO_6}$ and more generally illustrate the importance of complementing bulk dc magnetization measurements with nanoscale probes of spin ordering and spin fluctuations, as these nanoscale energetics and dynamics ultimately drive the functionality of emerging dissipationless spin-based information processing platforms.

\section{Methods}
We focus here on the high-$\mathrm{T_c}$ double perovskite $\mathrm{Sr_2FeReO_6}$, whose crystal structure is illustrated in Fig.~\ref{fig:fig1}(a). $\mathrm{Sr_2FeReO_6}$ is known for its robust ferromagnetism persistent up to 400K\cite{sohn2019room,zhang2022tunable}, and unlike most metallic ferromagnets, the ferromagnetism in $\mathrm{Sr_2FeReO_6}$ is attributed to a hybridization mechanism where the FM order is stabilized  by significant exchange splitting of ions and spin-dependent hybridization. A high quality epitaxial thin film of $\mathrm{Sr_2FeReO_6}$ with thickness $\sim 35$ nm was grown with a sintered $\mathrm{Sr_2FeReO_6}$ target by pulsed laser deposition on a (001) $\mathrm{SrTiO_3}$ substrate as previously described\cite{sohn2019room}.

Figure \ref{fig:fig1}(a) illustrates the scanning NV experiment conducted here using a Qnami ProteusQ microscope, where single-NV spin states are manipulated via optical and radio frequency (RF) pulses. The NV spin is pumped into the $| 0 \rangle$ state by an off-resonant 520 nm optical excitation, an RF excitation with frequency around 3 GHz is used to manipulate the ground state spin, and the NV spin can be read out by measuring the NV luminescence intensity, which is larger for the $| 0 \rangle$ state than the Zeeman-split $| \pm 1 \rangle$ states \cite{casola2018probing}.
The probe used here incorporates a single negatively-charged NV center implanted $\sim 10$ nm below the surface of a (100) diamond nanopillar integrated into a tuning-fork AFM probe. The NV center axis is orientated at an angle of about $55^{\circ}$ relative to the surface normal and is thus sensitive to changes in both in-plane and out-of-plane fields.

\begin{figure*}[ht]
\centering
    \includegraphics[width=6.5in]{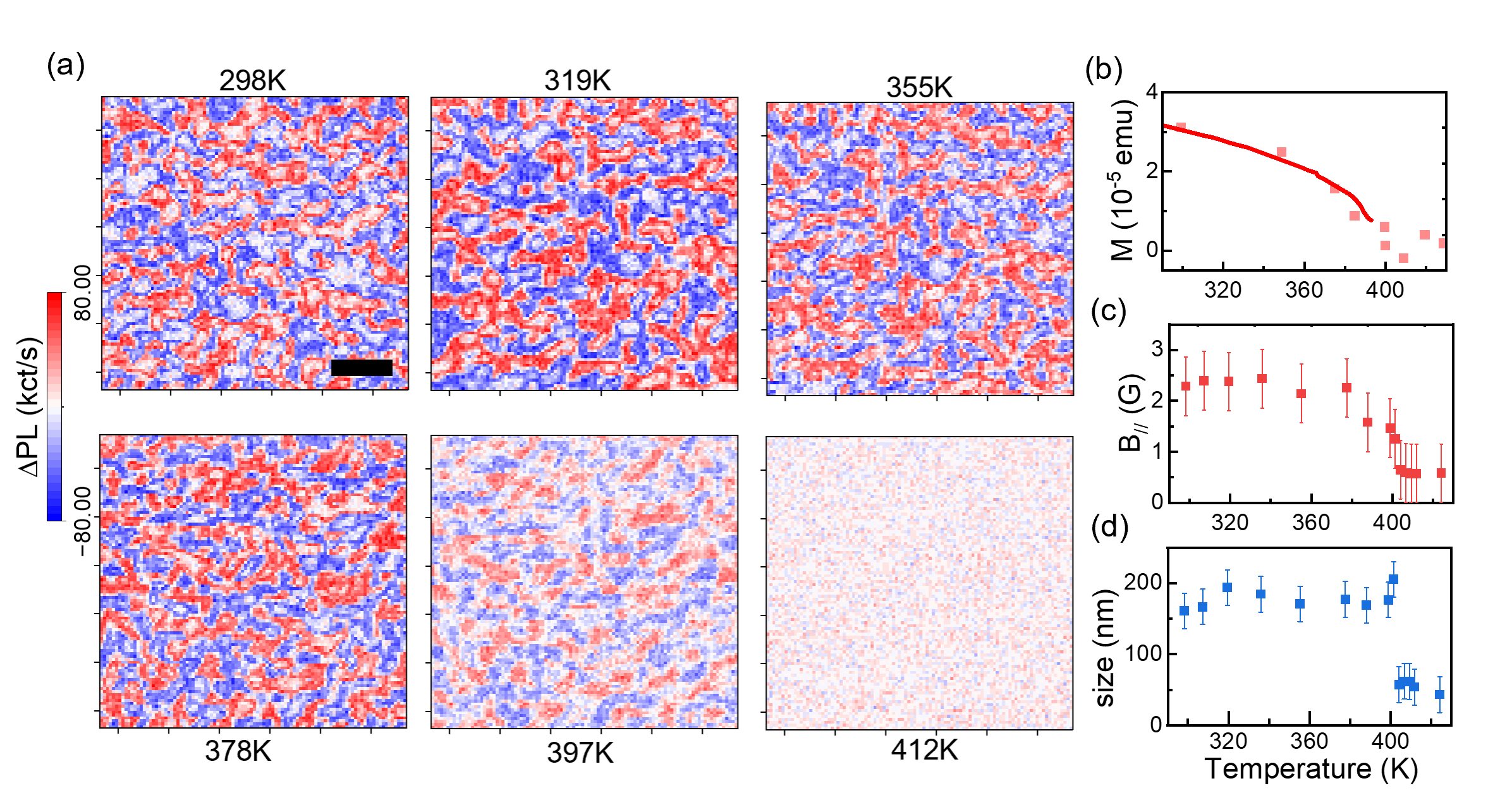}
    \captionsetup{justification=raggedright}
    \caption{ (a) Dual-iso-B scans acquired as a function of temperature for representative temperatures between 298 K and 412K. Scale bar: 1$\mu$m. (b) In-plane bulk M(T) curve of $\mathrm{Sr_2FeReO_6}$ measured at $H=1$ kOe. Solid dots are $\mathrm{M_S(T)}$ data obtained using a high temperature oven setup to determine the $\mathrm{T_c}$. Isothermal magnetization curves were measured between $\pm$4 kOe due to sample holder size limitations. The data above $\pm$1 kOe were linearly fit to estimate $\mathrm{M_S(T)}$.(c) NV sensed magnetic field strength. (d) Magnetic texture size estimated from temperature dependent NV dual-iso-B magnetometry scans. 
}
    \label{fig:fig2}
\end{figure*}

This approach enables nanoscale resolution, achieving detail down to 50 nm, by mapping surface magnetic textures through ODMR measurements taken at each pixel of the image. In full-B mode, the magnetic field is calculated from the frequency separation between the $| \pm 1 \rangle$ states observed in the ODMR spectrum, as shown in the left of Fig.~\ref{fig:fig1}(c) via the relationship \(B_z=\frac{\Delta\nu}{2\gamma}\) where \(\Delta\nu\) represents the frequency separation and \(\gamma=28GHz/T\) is the electron gyromagnetic ratio. However, acquiring a full ODMR spectrum at each pixel of an image can be time consuming; the dual-iso-B measurement protocol provides an alternative high-speed measurement modality in which a single ODMR spectrum is acquired in order to define the frequencies $f_1$ and $f_2$ illustrated in the right of Fig. \ref{fig:fig1}(c) at the full-width half-maximum of a $| 0 \rangle$ to $| \pm 1 \rangle$ transition. The luminescence intensity is then measured at each pixel of the image for RF excitation at $f_1$ and $f_2$. For small changes in magnetic field, the differential photoluminescence intensity $\Delta PL=PL(f_1)-PL(f_2)$ can be mapped to changes in the magnetic field semi-quantitatively.

Figure \ref{fig:fig1}(d) displays the nanoscale magnetic surface texture of a ferromagnetic $\mathrm{Sr_2FeReO_6}$ thin film measured in full-B mode (left) and dual-iso-B mode (right). The resolved magnetic domains exhibit structure across measured length scales spanning $\sim 50$ nm to 15 $\mu$m (with a 2 $\mu$m illustrated field of view in Figure \ref{fig:fig1}(d)),and the measured magnetic field varies by 10-15 Gauss across the full-B map. This dynamic range is sufficiently small to allow for accurate high-speed measurements to be acquired in dual-iso-B mode, as is seen in the qualitative one-to-one relationship between the full-B and dual-iso-B maps in Fig.~\ref{fig:fig1}(d).

\section{Result and Discussion}
The nanoscale resolution of magnetic textures under external stimuli, such as magnetic field and temperature, provides valuable insights into magnetic ordering in correlated materials that are critical to the development of dissipationless electronic and spintronic devices, as bulk magnetization measurements do not provide a complete picture of spin interactions within and between domains. In the absence of a strong magnetic field, the thermally induced FM-PM phase transition exhibits characteristics of a second-order phase transition\cite{Yoshioka2007,zinn2007phase,nishimori2011elements}. We probed this transition with temperature-dependent scanning NV microscopy, which was performed in dual-iso-B mode to minimize deleterious effects from drift at elevated temperatures. In Fig.~\ref{fig:fig2}(a), the FM-PM transition is clearly visualized between 397K and 412K with the disappearance of any measurable long-range magnetic order. To quantify this behavior, we performed a contrast analysis by comparing the most and least intense areas (top and bottom 10\% average, respectively) from the distribution in each false-color map, and we extracted the effective magnetic field strength using an inverse Lorentzian transformation linking count differences to field-induced ODMR shifts. As illustrated in Fig.~\ref{fig:fig2}(c), the stray magnetic field strength, and the associated surface magnetization rapidly drops as the temperature approaches 400K. We further compare the extracted magnetic field strength with bulk magnetization measurement data acquired with a superconducting quantum interference device\cite{sohn2019room} in Fig.~\ref{fig:fig2}(b). The data in Figs.~\ref{fig:fig2}(b) and \ref{fig:fig2}(c) show excellent agreement in the onset and gradual increase of the magnetization near the phase transition. 

Further analysis involved binarizing the images and performing auto-correlations to estimate the average spatial length scale of the magnetic texture (see Supporting Information). Interestingly, this analysis reveals that the average texture size remains constant ($\sim 200$ nm) over a broad temperature range from 298K to 400K [see Fig.~\ref{fig:fig2}(d)]. We note that this magnetic texture evolution contrasts with first-order transitions driven by external fields where domains gradually grow as long-range magnetic order is established with defects and local disorder serving as nucleation sites.

\begin{figure}[t]
\centering
\includegraphics[width=300pt]{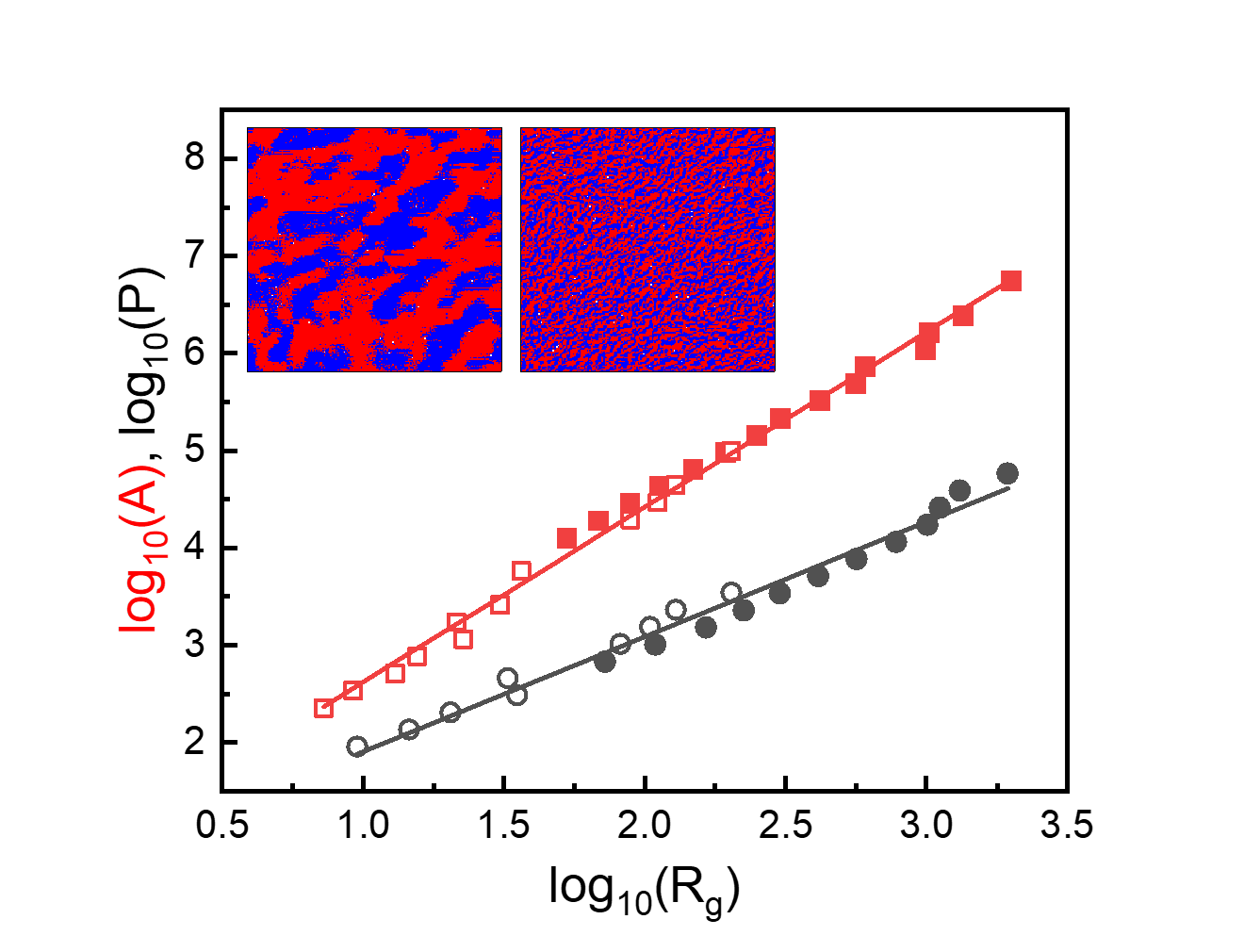}
\captionsetup{justification=raggedright}
\caption{Scaling analysis of cluster area ($A$) and perimeter ($P$) against the gyration radius \(R_g\) from dual-iso-B maps acquired in the ordered phase ($\mathrm{T<T_c}$). The plotted data points are derived by combining data from the maps shown in the inset that were each acquired with $200 \times 200$ pixel resolution across (left) 2 $\mu$m and (right) 15 $\mu$m fields of view.
}
    \label{fig:fig3}
\end{figure}

The gradual onset of magnetization, coupled with the observed texture evolution, is characteristic of a second-order phase transition, where power law scaling behavior is anticipated. To evaluate this scaling in the measured magnetic domain structure, we extract the radius of gyration (\(R_g\)), a parameter that represents the average distance between two points within a given magnetic cluster. Near the phase transition, NV magnetometry scans of the ordered phase $(T<T_c)$ were performed with both large (15 $\mu$m) and small (2 $\mu$m) fields of view. The bi-logarithmic plot of \(R_g\) against cluster area ($A$) shown in Fig. \ref{fig:fig3} reveals a power law scaling behavior spanning two and half decades, with the exponent $d_v=1.73 \pm 0.04$. Similarly, a robust power-law scaling is observed in the cluster perimeter ($P$) as a function of \(R_g\), with the exponent  $d_h=1.34 \pm 0.04$. Critical phenomena are governed by the concept of universality, where critical exponents depend on dimensionality and symmetry rather than material-specific details. In our result where SFRO is probed near the critical point ($T_c$), these critical exponents suggest a three-dimensional universality class for the spin system (see Supporting Information).  Characterizing critical exponents is crucial and our results highlight the power of NV magnetometry in understanding critical phenomena in second-order magnetic phase transitions.

On the other hand, magnetic fluctuations, which reflect dynamical deviations from spin equilibrium, can provide a complementary understanding of critical phenomena near continuous phase transitions. Measurements of the NV spin relaxation lifetime $T_1$ and coherence time $T_2$ have increasingly emerged as sensitive quantum probes of magnetic noise in an extensive frequency range.\cite{kumar2024room,rollo2021quantitative,lamichhane2024nitrogen,mclaughlin2022quantum}  In particular, the change in $T_1$ relative to the intrinsic $T_1$ for an NV center interacting with a fluctuating spin system via dipole-dipole interactions is proportional to the dynamical spin structure factor at the intrinsic NV frequency corresponding to the level splitting.

\begin{figure*}[ht]
\centering
    \includegraphics[width=6.5in]{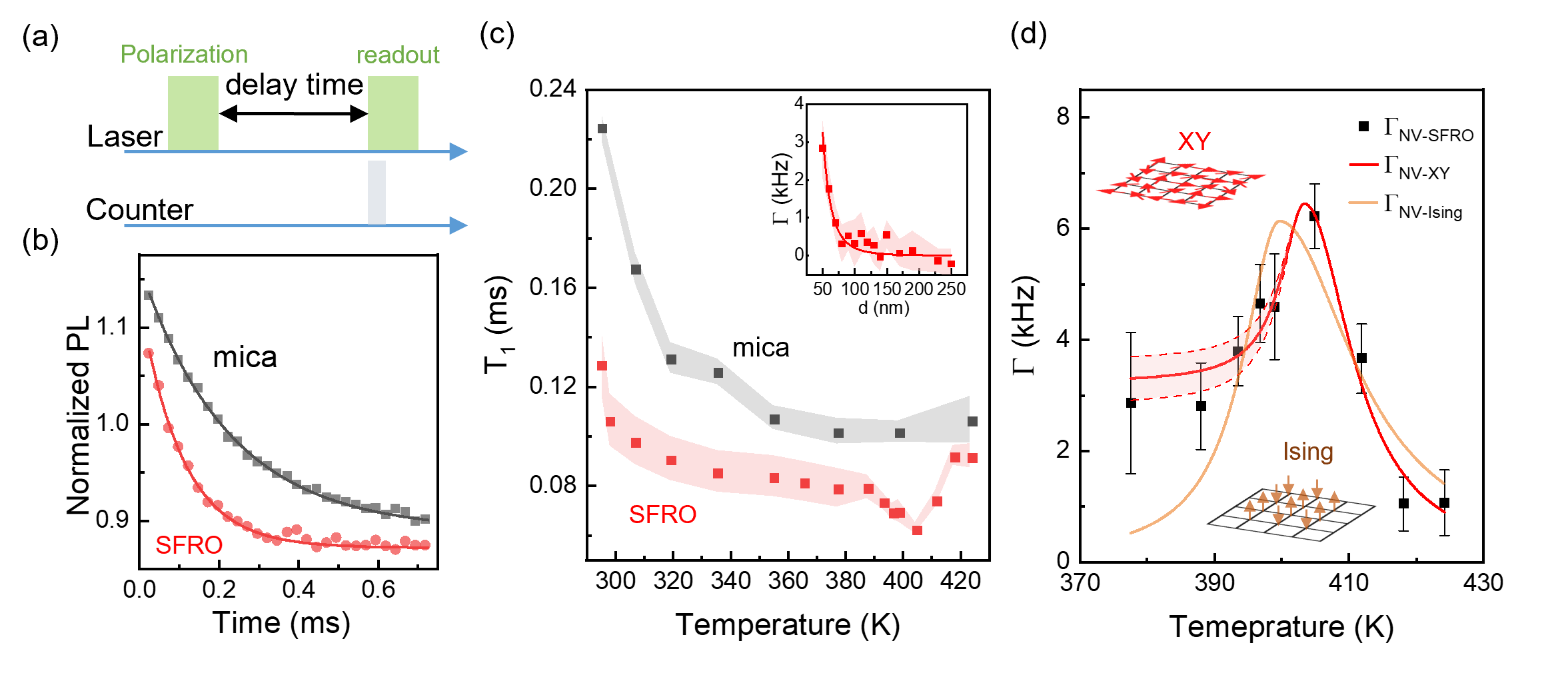}
    \captionsetup{justification=raggedright}
    \caption{(a) Pulse sequence scheme for all-optical NV $T_1$ relaxometry. (b) NV $T_1$ relaxation when a single NV is brought into proximity of $\mathrm{Sr_2FeReO_6}$ (red) and mica (black). (c) Temperature dependent $T_1$ spin relaxation time of a NV center interacting with $\mathrm{Sr_2FeReO_6}$ (red) and mica (black). The uncertainty (shaded area) is calculated based on a procedure described in the Supporting Information. Inset: Height-dependent relaxation rate \(\Gamma\) is calculated as $\Gamma_{\mathrm{SFRO}}-\Gamma_{\mathrm{mica}}$ at 405 K and fitted to the functional form $\frac{1} {d^3} - \frac{1} {(d+D)^3}$. (d) NV relaxation rate ($\Gamma$) near the FM-PM second order phase transition. The experimental data (black dots) are fitted with Landau-Ginzburg theories assuming XY and Ising universality classes. In the XY case, the solid line corresponds to an in-plane NV rotation angle of $\theta = 45^\circ$, with the shaded region representing the possible range of $\theta$ between 0$^\circ$ and 90$^\circ$ (see Supporting Information for details).
}
    \label{fig:fig4}
\end{figure*}

In the all-optical $T_1$ relaxometry sensing protocol used here, the NV spin state is initially polarized to $m_s=0$ by an optical pulse. The relaxation process is then monitored by a second optical pulse reading out the luminescence intensity as a function of delay time, as depicted in Fig.~\ref{fig:fig4}(a). A similar NV $T_1$ relaxometry technique has recently been demonstrated to capture the increased magnetic noise near a magnetic phase transition in 2D $\alpha$-Fe$_2$O$_3$ \cite{wang2022noninvasive}. In that context, the increase in magnetic fluctuations inferred from the $T_1$ relaxation is relevant to the in-plane/out-of-plane switching of antiferromagnetic ordering. To study magnetic ordering and the FM-PM continuous phase transition, we first compare $T_1$ relaxation times for the NV tip brought into proximity with $\mathrm{Sr_2FeReO_6}$ and mica (respectively) for two reasons. First, the spin relaxation time of NV centers can vary under different tip conditions, as NV centers are known to be sensitive to surface condition and nearby impurities\cite{de2020temperature,kumar2024room}. Second, in diamagnetic systems like mica, the magnetic dipole-dipole interaction is absent due to the diminished net magnetic moment of paired electrons. Therefore, the NV $T_1$ on mica serves as a reference for the intrinsic temperature-dependent NV spin dynamics. Representative $T_1$ measurements acquired at 300K on mica and $\mathrm{Sr_2FeReO_6}$ are shown in Fig.~\ref{fig:fig4}(b), and the temperature-dependent spin lifetime is shown in Fig.~\ref{fig:fig4}(c).

Below $T_c$, the NV center $T_1$ is suppressed for ferromagnetic $\mathrm{Sr_2FeReO_6}$ compared to the baseline temperature-dependent NV $T_1$ acquired on mica as a result of magnetic noise generated by the ferromagnetic spin bath. According to the fluctuation-dissipation theorem, the magnetic noise of a system is related to the imaginary part of the dynamical magnetic susceptibility \cite{casola2018probing,wang2022noninvasive,khoo2022probing}. At temperatures approaching $T_c=405$ K, Figure \ref{fig:fig4}(c) highlights the additional increase in the NV center spin relaxation rate compared with the intrinsic spin dynamics measured on the mica substrate, which can be understood as a direct result of the magnetic susceptibility diverging near a second order phase transition.\cite{Yoshioka2007, zinn2007phase,ma2014phase} Near $T_c$, the system is in a state of critical instability where the spins become increasingly correlated over large distances.\cite{jin2020imaging,nishimori2011elements} This increased spin-spin correlation leads to enhanced fluctuations as the spins spontaneously align with perturbing fields. NV relaxometry, given its susceptibility to magnetic fluctuations, is therefore a powerful tool for the characterization of continuous magnetic phase transitions.

To provide a more quantitative understanding, we apply a mean-field Landau–Ginzburg theory, which effectively captures the magnetic fluctuations near $T_c$. This theory incorporates the interaction between a NV center and a spin bath based on different spin models. As described in greater detail in the Supporting Information, we treat $\mathrm{Sr_2FeReO_6}$ as an infinite slab of thickness $D$ with magnetic fluctuations measured by a NV center at distance $d$ above its surface. For the Ising universality class, the NV relaxation rate takes the general form
\begin{equation}
\frac{1} {T_1} \propto T \left[ \frac{1} {d^3} - \frac{1} {(d+D)^3} \right] \frac{16} {\alpha^2 \max (\tau, -2\tau)^2 + \gamma^2 \omega_0^2}, \label{eq-ising-T1}
\end{equation}
where $\tau = T - T_c$ is the relative temperature, $\omega_0$ is the intrinsic frequency of the NV center, $\gamma$ is a damping coefficient, and $\alpha$ is an unknown Landau parameter. In the XY universality class, the same relaxation rate becomes
\begin{equation}
\frac{1} {T_1} \propto T \left[ \frac{1} {d^3} - \frac{1} {(d+D)^3} \right] \Bigg\{ \frac{7 + 2 \sin^2 \theta} {\alpha^2 \max (\tau, -2\tau)^2 + \gamma^2 \omega_0^2} + \frac{7 + 2 \cos^2 \theta} {\alpha^2 \max (\tau, 0)^2 + \gamma^2 \omega_0^2} \Bigg\}, \label{eq-xy-T1}
\end{equation}
where $\theta$ is an in-plane rotation angle between the NV direction and the given ferromagnetic domain. (Note that the relaxation rate in Eq.~(\ref{eq-xy-T1}) becomes independent of $\theta$ in the paramagnetic phase, $\tau>0$, as expected.) Unlike in the Ising universality class, the NV relaxation rate in the XY universality class remains finite in the ferromagnetic phase even for large relative temperatures, $|\tau| \gg \gamma \omega_0 / \alpha$, which is consistent with the measured temperature-dependent spin dynamics plotted in Fig. \ref{fig:fig4}(d). Indeed, while the height-dependent spin relaxation rate plotted in the inset of Fig.~\ref{fig:fig4}(c) is consistent with both Ising and XY universality classes due to the same $\frac{1}{d^3}-\frac{1}{(d+D)^3}$ scaling, the temperature-dependent behavior in the ferromagnetic phase [see Fig.~\ref{fig:fig4}(d)] points to the XY universality class. 

%The height dependence of $1/T_1$ registers into the LG model with a factor of q3e-2qd from the dipolar interaction. We note the algebraic scaling of height-dependent $1/T_1$ reveals fundamental properties of the interrogated magnetic systems that interacting with NV centers, as it is theoretically proposed for detection of collective excitation and magnetic ordering of quantum spin systems. In our experiment (near phase transition, 405K), however, with the estimated NV flying height (d~50nm) being larger than the film thickness (D~30-40nm), the height dependence could effectively factors out and scales as \(frac{1}{d^3}-frac{1}{(d+D)^3}\). Lending support to the model, the $T_1$ relaxation rate as a function of NV flying height is well fitted with the trend, as depicted in the inset of Fig. 4(c). 

To further examine the magnetic texture dependent $T_1$, we conducted an iso-$T_1$ scan where the decay in fluorescence intensity is monitored at a given delay time as a function of NV center position (see Supporting Information). The $T_1$ time for an NV center interacting with local magnetic textures can be estimated assuming a single exponential decay behavior across the scanning area. One might naively expect $T_1$ to vary with the spatially varying magnetic texture as a result of reduced cross-relaxation \cite{jarmola2012temperature,mrozek2015longitudinal} as the magnetic field lifts the spin degeneracy. However, no statistically significant correspondence was found between $T_1$ and the mapped magnetic texture. This discrepancy may arise from the complex magnetic structure of $\mathrm{Sr_2FeReO_6}$, which exhibits magnetic anisotropy between in-plane and out-of-plane directions, as indicated by bulk magnetization measurements (see Supporting Information). This complexity hinders a straightforward correlation between stray field profiles and magnetic domain morphology, thus complicating the comparison of spin noise contributions from domain walls and domains. Spatially-resolved $T_1$ measurements were also examined at elevated temperatures, including $T \sim T_c$, where the iso-$T_1$ maps showed no apparent spatial dependence on weak magnetic textures.

%We further formulate the NV relaxation rate of NV interacting with different spin models belonging to distinct universality classes (see detail in Supporting Information). In the Ising universality class, the corresponding Green function can be generally expressed as .... . Conversely, the green function for the XY model, accounting for non-zero in-plane magnetization components of $\mathrm{Sr_2FeReO_6}$, is given by.... Figure 4(d) highlights the experiment data where the enhance $T_1$ relaxation rate is calculated as \(Sigma_{SFRO}-Sigma_{mica}\). The experiment data near phase transition is fitted with modeling of NV-XY and NV-Ising interactions. We note that the NV-XY model captures the essential characteristic of the experiment observations with better match to Tc together with a pronounced and sustained magnetic fluctuation at temperature T<Tc (ferromagnetic ordering). We thus argue the ferromagnetic $\mathrm{Sr_2FeReO_6}$ thin film can be modeled as 3D XY-spin model. We note a dimension crossover might happen with film thickness further reduced to the length scale of spin-spin interaction within the system. The universality of spin model in this scenario thus pose a interesting question as there is no long range magnetic order supported with non-Ising model in 2D limit. 

In conclusion, we  have studied the continuous phase transition of the high-$T_c$ ferromagnetic oxide $\mathrm{Sr_2FeReO_6}$ using scanning NV magnetometry and relaxometry, which provide quantitative, minimally perturbative probes of spin ordering and spin dynamics. The power-law scaling behavior observed in the magnetic texture is a powerful way of understanding the phase transition and suggests that the critical point is in a 3D universality class. With the magnetic fluctuations complementarily sensed by NV relaxometry across the phase transition, we can then argue that the ferromagnetic $\mathrm{Sr_2FeReO_6}$ thin film can be modeled as a 3D XY spin system on length scales comparable to the film thickness. We note that a dimensional crossover might happen if the film thickness is further reduced to the length scale of the spin-spin interactions. The nature of the phase transition in this scenario would pose an interesting question as there is no long range magnetic order for spin systems with continuous symmetries in the 2D limit.

%We note that, on theoretical front, the understanding of continuous phase transition and calculating critical exponents in various models have proved to be a challenge. Our work highlights the scanning NV capability with non-invasive resolution of nanoscale magnetic texture and its dynamical sensitivity of fluctuations, essential characteristics of continuous phase transition.
\begin{tocentry}
\centering
\includegraphics[width=\textwidth]{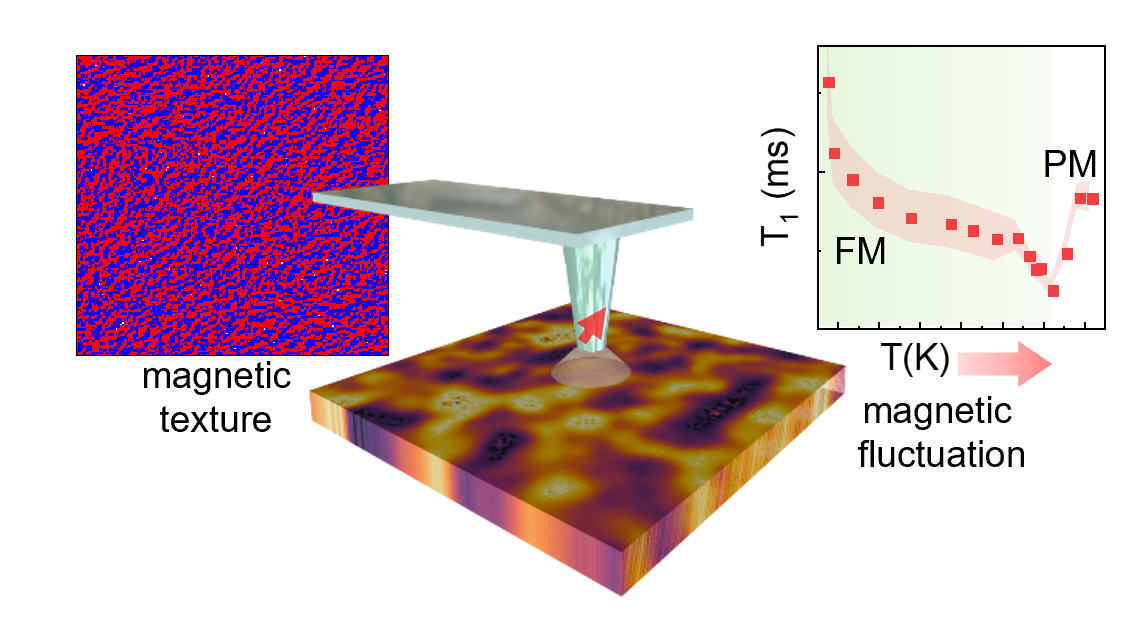} % Your TOC figure
\textbf{Table of Contents Graphic} % Use bold text or no caption to avoid issues
\end{tocentry}

\begin{acknowledgement}
This research was sponsored by the U. S. Department of Energy, Office of Science, Basic Energy Sciences, Materials Sciences and Engineering Division and in part by the Computational Materials Sciences Program and Center for Predictive Simulation of
Functional Materials.. Scanning NV microscopy was supported by the Center for Nanophase Materials Sciences (CNMS), which is a US Department of Energy, Office of Science User Facility at Oak Ridge National Laboratory. 
YS and EC acknowledge support from from the Department of Energy under grant DOE-QIS (DE-FOA-0002449). 
EC acknowledges support from NSF Grant no. DMR-2006192. KD thanks the University of Illinois at Urbana Champaign for support.
\end{acknowledgement}

%\bibliography{references}

\begin{mcitethebibliography}{35}
\providecommand*\natexlab[1]{#1}
\providecommand*\mciteSetBstSublistMode[1]{}
\providecommand*\mciteSetBstMaxWidthForm[2]{}
\providecommand*\mciteBstWouldAddEndPuncttrue
  {\def\EndOfBibitem{\unskip.}}
\providecommand*\mciteBstWouldAddEndPunctfalse
  {\let\EndOfBibitem\relax}
\providecommand*\mciteSetBstMidEndSepPunct[3]{}
\providecommand*\mciteSetBstSublistLabelBeginEnd[3]{}
\providecommand*\EndOfBibitem{}
\mciteSetBstSublistMode{f}
\mciteSetBstMaxWidthForm{subitem}{(\alph{mcitesubitemcount})}
\mciteSetBstSublistLabelBeginEnd
  {\mcitemaxwidthsubitemform\space}
  {\relax}
  {\relax}

\bibitem[\v{Z}aper \latin{et~al.}(2024)\v{Z}aper, Rickhaus, Wyss, Gross,
  Wagner, Poggio, and Braakman]{zaper2024scanning}
\v{Z}aper,~L.; Rickhaus,~P.; Wyss,~M.; Gross,~B.; Wagner,~K.; Poggio,~M.;
  Braakman,~F. Scanning Nitrogen-Vacancy Magnetometry of
  Focused-Electron-Beam-Deposited Cobalt Nanomagnets. \emph{ACS Applied Nano
  Materials} \textbf{2024}, \relax
\mciteBstWouldAddEndPunctfalse
\mciteSetBstMidEndSepPunct{\mcitedefaultmidpunct}
{}{\mcitedefaultseppunct}\relax
\EndOfBibitem
\bibitem[Celano \latin{et~al.}(2021)Celano, Zhong, Ciubotaru, Stoleriu, Stark,
  Rickhaus, de~Oliveira, Munsch, Favia, Korytov, \latin{et~al.}
  others]{celano2021probing}
Celano,~U.; Zhong,~H.; Ciubotaru,~F.; Stoleriu,~L.; Stark,~A.; Rickhaus,~P.;
  de~Oliveira,~F.~F.; Munsch,~M.; Favia,~P.; Korytov,~M., \latin{et~al.}
  Probing magnetic defects in ultra-scaled nanowires with optically detected
  spin resonance in nitrogen-vacancy center in diamond. \emph{Nano Letters}
  \textbf{2021}, \emph{21}, 10409--10415\relax
\mciteBstWouldAddEndPuncttrue
\mciteSetBstMidEndSepPunct{\mcitedefaultmidpunct}
{\mcitedefaultendpunct}{\mcitedefaultseppunct}\relax
\EndOfBibitem
\bibitem[Sun \latin{et~al.}(2021)Sun, Song, Anderson, Brunner, F{\"o}rster,
  Shalomayeva, Taniguchi, Watanabe, Gr{\"a}fe, St{\"o}hr, \latin{et~al.}
  others]{sun2021magnetic}
Sun,~Q.-C.; Song,~T.; Anderson,~E.; Brunner,~A.; F{\"o}rster,~J.;
  Shalomayeva,~T.; Taniguchi,~T.; Watanabe,~K.; Gr{\"a}fe,~J.; St{\"o}hr,~R.,
  \latin{et~al.}  Magnetic domains and domain wall pinning in atomically thin
  $\mathrm{CrBr_3}$ revealed by nanoscale imaging. \emph{Nature Communications}
  \textbf{2021}, \emph{12}, 1989\relax
\mciteBstWouldAddEndPuncttrue
\mciteSetBstMidEndSepPunct{\mcitedefaultmidpunct}
{\mcitedefaultendpunct}{\mcitedefaultseppunct}\relax
\EndOfBibitem
\bibitem[Meisenheimer \latin{et~al.}(2024)Meisenheimer, Moore, Zhou, Zhang,
  Huang, Husain, Chen, Martin, Persson, Griffin, \latin{et~al.}
  others]{meisenheimer2024switching}
Meisenheimer,~P.; Moore,~G.; Zhou,~S.; Zhang,~H.; Huang,~X.; Husain,~S.;
  Chen,~X.; Martin,~L.~W.; Persson,~K.~A.; Griffin,~S., \latin{et~al.}
  Switching the spin cycloid in BiFeO3 with an electric field. \emph{Nature
  Communications} \textbf{2024}, \emph{15}, 2903\relax
\mciteBstWouldAddEndPuncttrue
\mciteSetBstMidEndSepPunct{\mcitedefaultmidpunct}
{\mcitedefaultendpunct}{\mcitedefaultseppunct}\relax
\EndOfBibitem
\bibitem[Ding \latin{et~al.}(2023)Ding, Sun, Zheng, Ma, Wang, Zang, Yu, Chen,
  Wang, Wang, \latin{et~al.} others]{ding2023observation}
Ding,~Z.; Sun,~Y.; Zheng,~N.; Ma,~X.; Wang,~M.; Zang,~Y.; Yu,~P.; Chen,~Z.;
  Wang,~P.; Wang,~Y., \latin{et~al.}  Observation of uniaxial strain tuned spin
  cycloid in a freestanding BiFeO3 film. \emph{Advanced Functional Materials}
  \textbf{2023}, \emph{33}, 2213725\relax
\mciteBstWouldAddEndPuncttrue
\mciteSetBstMidEndSepPunct{\mcitedefaultmidpunct}
{\mcitedefaultendpunct}{\mcitedefaultseppunct}\relax
\EndOfBibitem
\bibitem[Jenkins \latin{et~al.}(2019)Jenkins, Pelliccione, Yu, Ma, Li, Wang,
  and Jayich]{jenkins2019single}
Jenkins,~A.; Pelliccione,~M.; Yu,~G.; Ma,~X.; Li,~X.; Wang,~K.~L.; Jayich,~A.
  C.~B. Single-spin sensing of domain-wall structure and dynamics in a
  thin-film skyrmion host. \emph{Physical Review Materials} \textbf{2019},
  \emph{3}, 083801\relax
\mciteBstWouldAddEndPuncttrue
\mciteSetBstMidEndSepPunct{\mcitedefaultmidpunct}
{\mcitedefaultendpunct}{\mcitedefaultseppunct}\relax
\EndOfBibitem
\bibitem[Sch{\"a}fer-Nolte \latin{et~al.}(2014)Sch{\"a}fer-Nolte, Schlipf,
  Ternes, Reinhard, Kern, and Wrachtrup]{schafer2014tracking}
Sch{\"a}fer-Nolte,~E.; Schlipf,~L.; Ternes,~M.; Reinhard,~F.; Kern,~K.;
  Wrachtrup,~J. Tracking temperature-dependent relaxation times of ferritin
  nanomagnets with a wideband quantum spectrometer. \emph{Physical Review
  Letters} \textbf{2014}, \emph{113}, 217204\relax
\mciteBstWouldAddEndPuncttrue
\mciteSetBstMidEndSepPunct{\mcitedefaultmidpunct}
{\mcitedefaultendpunct}{\mcitedefaultseppunct}\relax
\EndOfBibitem
\bibitem[McCullian \latin{et~al.}(2020)McCullian, Thabt, Gray, Melendez, Wolf,
  Safonov, Pelekhov, Bhallamudi, Page, and Hammel]{mccullian2020broadband}
McCullian,~B.~A.; Thabt,~A.~M.; Gray,~B.~A.; Melendez,~A.~L.; Wolf,~M.~S.;
  Safonov,~V.~L.; Pelekhov,~D.~V.; Bhallamudi,~V.~P.; Page,~M.~R.;
  Hammel,~P.~C. Broadband multi-magnon relaxometry using a quantum spin sensor
  for high frequency ferromagnetic dynamics sensing. \emph{Nature
  Communications} \textbf{2020}, \emph{11}, 5229\relax
\mciteBstWouldAddEndPuncttrue
\mciteSetBstMidEndSepPunct{\mcitedefaultmidpunct}
{\mcitedefaultendpunct}{\mcitedefaultseppunct}\relax
\EndOfBibitem
\bibitem[Du \latin{et~al.}(2017)Du, Van~der Sar, Zhou, Upadhyaya, Casola,
  Zhang, Onbasli, Ross, Walsworth, Tserkovnyak, \latin{et~al.}
  others]{du2017control}
Du,~C.; Van~der Sar,~T.; Zhou,~T.~X.; Upadhyaya,~P.; Casola,~F.; Zhang,~H.;
  Onbasli,~M.~C.; Ross,~C.~A.; Walsworth,~R.~L.; Tserkovnyak,~Y.,
  \latin{et~al.}  Control and local measurement of the spin chemical potential
  in a magnetic insulator. \emph{Science} \textbf{2017}, \emph{357},
  195--198\relax
\mciteBstWouldAddEndPuncttrue
\mciteSetBstMidEndSepPunct{\mcitedefaultmidpunct}
{\mcitedefaultendpunct}{\mcitedefaultseppunct}\relax
\EndOfBibitem
\bibitem[Finco \latin{et~al.}(2021)Finco, Haykal, Tanos, Fabre, Chouaieb,
  Akhtar, Robert-Philip, Legrand, Ajejas, Bouzehouane, \latin{et~al.}
  others]{finco2021imaging}
Finco,~A.; Haykal,~A.; Tanos,~R.; Fabre,~F.; Chouaieb,~S.; Akhtar,~W.;
  Robert-Philip,~I.; Legrand,~W.; Ajejas,~F.; Bouzehouane,~K., \latin{et~al.}
  Imaging non-collinear antiferromagnetic textures via single spin relaxometry.
  \emph{Nature communications} \textbf{2021}, \emph{12}, 767\relax
\mciteBstWouldAddEndPuncttrue
\mciteSetBstMidEndSepPunct{\mcitedefaultmidpunct}
{\mcitedefaultendpunct}{\mcitedefaultseppunct}\relax
\EndOfBibitem
\bibitem[Kolkowitz \latin{et~al.}(2015)Kolkowitz, Safira, High, Devlin, Choi,
  Unterreithmeier, Patterson, Zibrov, Manucharyan, Park, \latin{et~al.}
  others]{kolkowitz2015probing}
Kolkowitz,~S.; Safira,~A.; High,~A.; Devlin,~R.; Choi,~S.; Unterreithmeier,~Q.;
  Patterson,~D.; Zibrov,~A.; Manucharyan,~V.; Park,~H., \latin{et~al.}  Probing
  Johnson noise and ballistic transport in normal metals with a single-spin
  qubit. \emph{Science} \textbf{2015}, \emph{347}, 1129--1132\relax
\mciteBstWouldAddEndPuncttrue
\mciteSetBstMidEndSepPunct{\mcitedefaultmidpunct}
{\mcitedefaultendpunct}{\mcitedefaultseppunct}\relax
\EndOfBibitem
\bibitem[Andersen \latin{et~al.}(2019)Andersen, Dwyer, Sanchez-Yamagishi,
  Rodriguez-Nieva, Agarwal, Watanabe, Taniguchi, Demler, Kim, Park,
  \latin{et~al.} others]{andersen2019electron}
Andersen,~T.~I.; Dwyer,~B.~L.; Sanchez-Yamagishi,~J.~D.;
  Rodriguez-Nieva,~J.~F.; Agarwal,~K.; Watanabe,~K.; Taniguchi,~T.;
  Demler,~E.~A.; Kim,~P.; Park,~H., \latin{et~al.}  Electron-phonon instability
  in graphene revealed by global and local noise probes. \emph{Science}
  \textbf{2019}, \emph{364}, 154--157\relax
\mciteBstWouldAddEndPuncttrue
\mciteSetBstMidEndSepPunct{\mcitedefaultmidpunct}
{\mcitedefaultendpunct}{\mcitedefaultseppunct}\relax
\EndOfBibitem
\bibitem[McLaughlin \latin{et~al.}(2022)McLaughlin, Hu, Huang, Zhang, Lu, Yan,
  Wang, Tserkovnyak, Ni, and Du]{mclaughlin2022quantum}
McLaughlin,~N.~J.; Hu,~C.; Huang,~M.; Zhang,~S.; Lu,~H.; Yan,~G.~Q.; Wang,~H.;
  Tserkovnyak,~Y.; Ni,~N.; Du,~C.~R. Quantum imaging of magnetic phase
  transitions and spin fluctuations in intrinsic magnetic topological
  nanoflakes. \emph{Nano Letters} \textbf{2022}, \emph{22}, 5810--5817\relax
\mciteBstWouldAddEndPuncttrue
\mciteSetBstMidEndSepPunct{\mcitedefaultmidpunct}
{\mcitedefaultendpunct}{\mcitedefaultseppunct}\relax
\EndOfBibitem
\bibitem[Pelliccione \latin{et~al.}(2014)Pelliccione, Myers, Pascal, Das, and
  Jayich]{pelliccione2014two}
Pelliccione,~M.; Myers,~B.~A.; Pascal,~L.; Das,~A.; Jayich,~A.~B.
  Two-dimensional nanoscale imaging of gadolinium spins via scanning probe
  relaxometry with a single spin in diamond. \emph{Physical Review Applied}
  \textbf{2014}, \emph{2}, 054014\relax
\mciteBstWouldAddEndPuncttrue
\mciteSetBstMidEndSepPunct{\mcitedefaultmidpunct}
{\mcitedefaultendpunct}{\mcitedefaultseppunct}\relax
\EndOfBibitem
\bibitem[Schmid-Lorch \latin{et~al.}(2015)Schmid-Lorch, H\"aberle, Reinhard,
  Zappe, Slota, Bogani, Finkler, and Wrachtrup]{schmid2015relaxometry}
Schmid-Lorch,~D.; H\"aberle,~T.; Reinhard,~F.; Zappe,~A.; Slota,~M.;
  Bogani,~L.; Finkler,~A.; Wrachtrup,~J. Relaxometry and dephasing imaging of
  superparamagnetic magnetite nanoparticles using a single qubit. \emph{Nano
  Letters} \textbf{2015}, \emph{15}, 4942--4947\relax
\mciteBstWouldAddEndPuncttrue
\mciteSetBstMidEndSepPunct{\mcitedefaultmidpunct}
{\mcitedefaultendpunct}{\mcitedefaultseppunct}\relax
\EndOfBibitem
\bibitem[Tetienne \latin{et~al.}(2016)Tetienne, Lombard, Simpson, Ritchie, Lu,
  Mulvaney, and Hollenberg]{tetienne2016scanning}
Tetienne,~J.-P.; Lombard,~A.; Simpson,~D.~A.; Ritchie,~C.; Lu,~J.;
  Mulvaney,~P.; Hollenberg,~L.~C. Scanning nanospin ensemble microscope for
  nanoscale magnetic and thermal imaging. \emph{Nano Letters} \textbf{2016},
  \emph{16}, 326--333\relax
\mciteBstWouldAddEndPuncttrue
\mciteSetBstMidEndSepPunct{\mcitedefaultmidpunct}
{\mcitedefaultendpunct}{\mcitedefaultseppunct}\relax
\EndOfBibitem
\bibitem[McCullian(2020)]{mccullian2020detection}
McCullian,~B.~A. \emph{Detection of Ferromagnetic Dynamics Using NV Centers in
  Diamond}; The Ohio State University, 2020\relax
\mciteBstWouldAddEndPuncttrue
\mciteSetBstMidEndSepPunct{\mcitedefaultmidpunct}
{\mcitedefaultendpunct}{\mcitedefaultseppunct}\relax
\EndOfBibitem
\bibitem[Ziffer \latin{et~al.}(2024)Ziffer, Machado, Ursprung, Lozovoi, Tazi,
  Yuan, Ziebel, Delord, Zeng, Telford, \latin{et~al.}
  others]{ziffer2024quantum}
Ziffer,~M.~E.; Machado,~F.; Ursprung,~B.; Lozovoi,~A.; Tazi,~A.~B.; Yuan,~Z.;
  Ziebel,~M.~E.; Delord,~T.; Zeng,~N.; Telford,~E., \latin{et~al.}  Quantum
  Noise Spectroscopy of Critical Slowing Down in an Atomically Thin Magnet.
  \emph{arXiv preprint arXiv:2407.05614} \textbf{2024}, \relax
\mciteBstWouldAddEndPunctfalse
\mciteSetBstMidEndSepPunct{\mcitedefaultmidpunct}
{}{\mcitedefaultseppunct}\relax
\EndOfBibitem
\bibitem[Sohn \latin{et~al.}(2019)Sohn, Skoropata, Choi, Gao, Rastogi, Huon,
  McGuire, Nuckols, Zhang, Freeland, \latin{et~al.} others]{sohn2019room}
Sohn,~C.; Skoropata,~E.; Choi,~Y.; Gao,~X.; Rastogi,~A.; Huon,~A.;
  McGuire,~M.~A.; Nuckols,~L.; Zhang,~Y.; Freeland,~J.~W., \latin{et~al.}
  Room-Temperature Ferromagnetic Insulating State in Cation-Ordered
  Double-Perovskite $\mathrm{Sr_2Fe_{1+x}Re_{1-x}O_6}$ Films. \emph{Advanced
  Materials} \textbf{2019}, \emph{31}, 1805389\relax
\mciteBstWouldAddEndPuncttrue
\mciteSetBstMidEndSepPunct{\mcitedefaultmidpunct}
{\mcitedefaultendpunct}{\mcitedefaultseppunct}\relax
\EndOfBibitem
\bibitem[Zhang \latin{et~al.}(2022)Zhang, Yan, Huang, Chi, Li, Lim, Zeng, Han,
  Omar, Jin, \latin{et~al.} others]{zhang2022tunable}
Zhang,~Z.; Yan,~H.; Huang,~Z.; Chi,~X.; Li,~C.; Lim,~Z.~S.; Zeng,~S.; Han,~K.;
  Omar,~G.~J.; Jin,~K., \latin{et~al.}  Tunable Magnetic Properties in
  Sr2FeReO6 Double-Perovskite. \emph{Nano Letters} \textbf{2022}, \emph{22},
  9900--9906\relax
\mciteBstWouldAddEndPuncttrue
\mciteSetBstMidEndSepPunct{\mcitedefaultmidpunct}
{\mcitedefaultendpunct}{\mcitedefaultseppunct}\relax
\EndOfBibitem
\bibitem[Casola \latin{et~al.}(2018)Casola, Van Der~Sar, and
  Yacoby]{casola2018probing}
Casola,~F.; Van Der~Sar,~T.; Yacoby,~A. Probing condensed matter physics with
  magnetometry based on nitrogen-vacancy centres in diamond. \emph{Nature
  Reviews Materials} \textbf{2018}, \emph{3}, 1--13\relax
\mciteBstWouldAddEndPuncttrue
\mciteSetBstMidEndSepPunct{\mcitedefaultmidpunct}
{\mcitedefaultendpunct}{\mcitedefaultseppunct}\relax
\EndOfBibitem
\bibitem[Yos(2007)]{Yoshioka2007}
\emph{Statistical Physics: An Introduction}; Springer Berlin Heidelberg:
  Berlin, Heidelberg, 2007; pp 133--145\relax
\mciteBstWouldAddEndPuncttrue
\mciteSetBstMidEndSepPunct{\mcitedefaultmidpunct}
{\mcitedefaultendpunct}{\mcitedefaultseppunct}\relax
\EndOfBibitem
\bibitem[Zinn-Justin(2007)]{zinn2007phase}
Zinn-Justin,~J. \emph{Phase transitions and renormalization group}; Oxford
  University Press, 2007\relax
\mciteBstWouldAddEndPuncttrue
\mciteSetBstMidEndSepPunct{\mcitedefaultmidpunct}
{\mcitedefaultendpunct}{\mcitedefaultseppunct}\relax
\EndOfBibitem
\bibitem[Nishimori and Ortiz(2011)Nishimori, and Ortiz]{nishimori2011elements}
Nishimori,~H.; Ortiz,~G. \emph{Elements of phase transitions and critical
  phenomena}; Oxford university press, 2011\relax
\mciteBstWouldAddEndPuncttrue
\mciteSetBstMidEndSepPunct{\mcitedefaultmidpunct}
{\mcitedefaultendpunct}{\mcitedefaultseppunct}\relax
\EndOfBibitem
\bibitem[Kumar \latin{et~al.}(2024)Kumar, Yudilevich, Smooha, Zohar, Pariari,
  St\"{o}hr, Denisenko, H\"{u}cker, and Finkler]{kumar2024room}
Kumar,~J.; Yudilevich,~D.; Smooha,~A.; Zohar,~I.; Pariari,~A.~K.;
  St\"{o}hr,~R.; Denisenko,~A.; H\"{u}cker,~M.; Finkler,~A. Room Temperature
  Relaxometry of Single Nitrogen Vacancy Centers in Proximity to $\alpha$-RuCl3
  Nanoflakes. \emph{Nano Letters} \textbf{2024}, \emph{24}, 4793--4800\relax
\mciteBstWouldAddEndPuncttrue
\mciteSetBstMidEndSepPunct{\mcitedefaultmidpunct}
{\mcitedefaultendpunct}{\mcitedefaultseppunct}\relax
\EndOfBibitem
\bibitem[Rollo \latin{et~al.}(2021)Rollo, Finco, Tanos, Fabre, Devolder,
  Robert-Philip, and Jacques]{rollo2021quantitative}
Rollo,~M.; Finco,~A.; Tanos,~R.; Fabre,~F.; Devolder,~T.; Robert-Philip,~I.;
  Jacques,~V. Quantitative study of the response of a single NV defect in
  diamond to magnetic noise. \emph{Physical Review B} \textbf{2021},
  \emph{103}, 235418\relax
\mciteBstWouldAddEndPuncttrue
\mciteSetBstMidEndSepPunct{\mcitedefaultmidpunct}
{\mcitedefaultendpunct}{\mcitedefaultseppunct}\relax
\EndOfBibitem
\bibitem[Lamichhane \latin{et~al.}(2024)Lamichhane, Timalsina, Schultz,
  Fescenko, Ambal, Liou, Lai, and Laraoui]{lamichhane2024nitrogen}
Lamichhane,~S.; Timalsina,~R.; Schultz,~C.; Fescenko,~I.; Ambal,~K.;
  Liou,~S.-H.; Lai,~R.~Y.; Laraoui,~A. Nitrogen-Vacancy Magnetic Relaxometry of
  Nanoclustered Cytochrome C Proteins. \emph{Nano Letters} \textbf{2024},
  \emph{24}, 873--880\relax
\mciteBstWouldAddEndPuncttrue
\mciteSetBstMidEndSepPunct{\mcitedefaultmidpunct}
{\mcitedefaultendpunct}{\mcitedefaultseppunct}\relax
\EndOfBibitem
\bibitem[Wang \latin{et~al.}(2022)Wang, Zhang, McLaughlin, Flebus, Huang, Xiao,
  Liu, Wu, Fullerton, Tserkovnyak, \latin{et~al.} others]{wang2022noninvasive}
Wang,~H.; Zhang,~S.; McLaughlin,~N.~J.; Flebus,~B.; Huang,~M.; Xiao,~Y.;
  Liu,~C.; Wu,~M.; Fullerton,~E.~E.; Tserkovnyak,~Y., \latin{et~al.}
  Noninvasive measurements of spin transport properties of an antiferromagnetic
  insulator. \emph{Science Advances} \textbf{2022}, \emph{8}, eabg8562\relax
\mciteBstWouldAddEndPuncttrue
\mciteSetBstMidEndSepPunct{\mcitedefaultmidpunct}
{\mcitedefaultendpunct}{\mcitedefaultseppunct}\relax
\EndOfBibitem
\bibitem[de~Guillebon \latin{et~al.}(2020)de~Guillebon, Vindolet, Roch,
  Jacques, and Rondin]{de2020temperature}
de~Guillebon,~T.; Vindolet,~B.; Roch,~J.-F.; Jacques,~V.; Rondin,~L.
  Temperature dependence of the longitudinal spin relaxation time T 1 of single
  nitrogen-vacancy centers in nanodiamonds. \emph{Physical Review B}
  \textbf{2020}, \emph{102}, 165427\relax
\mciteBstWouldAddEndPuncttrue
\mciteSetBstMidEndSepPunct{\mcitedefaultmidpunct}
{\mcitedefaultendpunct}{\mcitedefaultseppunct}\relax
\EndOfBibitem
\bibitem[Khoo \latin{et~al.}(2022)Khoo, Pientka, Lee, and
  Villadiego]{khoo2022probing}
Khoo,~J.~Y.; Pientka,~F.; Lee,~P.~A.; Villadiego,~I.~S. Probing the quantum
  noise of the spinon Fermi surface with NV centers. \emph{Physical Review B}
  \textbf{2022}, \emph{106}, 115108\relax
\mciteBstWouldAddEndPuncttrue
\mciteSetBstMidEndSepPunct{\mcitedefaultmidpunct}
{\mcitedefaultendpunct}{\mcitedefaultseppunct}\relax
\EndOfBibitem
\bibitem[Ma and Wang(2014)Ma, and Wang]{ma2014phase}
Ma,~T.; Wang,~S. \emph{Phase transition dynamics}; Springer, 2014\relax
\mciteBstWouldAddEndPuncttrue
\mciteSetBstMidEndSepPunct{\mcitedefaultmidpunct}
{\mcitedefaultendpunct}{\mcitedefaultseppunct}\relax
\EndOfBibitem
\bibitem[Jin \latin{et~al.}(2020)Jin, Tao, Kang, Watanabe, Taniguchi, Mak, and
  Shan]{jin2020imaging}
Jin,~C.; Tao,~Z.; Kang,~K.; Watanabe,~K.; Taniguchi,~T.; Mak,~K.~F.; Shan,~J.
  Imaging and control of critical fluctuations in two-dimensional magnets.
  \emph{Nature Materials} \textbf{2020}, \emph{19}, 1290--1294\relax
\mciteBstWouldAddEndPuncttrue
\mciteSetBstMidEndSepPunct{\mcitedefaultmidpunct}
{\mcitedefaultendpunct}{\mcitedefaultseppunct}\relax
\EndOfBibitem
\bibitem[Jarmola \latin{et~al.}(2012)Jarmola, Acosta, Jensen, Chemerisov, and
  Budker]{jarmola2012temperature}
Jarmola,~A.; Acosta,~V.; Jensen,~K.; Chemerisov,~S.; Budker,~D. Temperature-and
  magnetic-field-dependent longitudinal spin relaxation in nitrogen-vacancy
  ensembles in diamond. \emph{Physical Review Letters} \textbf{2012},
  \emph{108}, 197601\relax
\mciteBstWouldAddEndPuncttrue
\mciteSetBstMidEndSepPunct{\mcitedefaultmidpunct}
{\mcitedefaultendpunct}{\mcitedefaultseppunct}\relax
\EndOfBibitem
\bibitem[Mr{\'o}zek \latin{et~al.}(2015)Mr{\'o}zek, Rudnicki, Kehayias,
  Jarmola, Budker, and Gawlik]{mrozek2015longitudinal}
Mr{\'o}zek,~M.; Rudnicki,~D.; Kehayias,~P.; Jarmola,~A.; Budker,~D.; Gawlik,~W.
  Longitudinal spin relaxation in nitrogen-vacancy ensembles in diamond.
  \emph{EPJ Quantum Technology} \textbf{2015}, \emph{2}, 1--11\relax
\mciteBstWouldAddEndPuncttrue
\mciteSetBstMidEndSepPunct{\mcitedefaultmidpunct}
{\mcitedefaultendpunct}{\mcitedefaultseppunct}\relax
\EndOfBibitem
\end{mcitethebibliography}
\providecommand{\latin}[1]{#1}
\makeatletter
\providecommand{\doi}
  {\begingroup\let\do\@makeother\dospecials
  \catcode`\{=1 \catcode`\}=2 \doi@aux}
\providecommand{\doi@aux}[1]{\endgroup\texttt{#1}}
\makeatother
\providecommand*\mcitethebibliography{\thebibliography}
\csname @ifundefined\endcsname{endmcitethebibliography}
  {\let\endmcitethebibliography\endthebibliography}{}

\end{document}